%% file: main.tex
\documentclass[%
 reprint,
 amsmath,amssymb,
 aps,
]{revtex4-2}

\usepackage{graphicx}
\usepackage{dcolumn}
\usepackage{bm}
\usepackage{microtype}

\usepackage{comment,mfirstuc}
\usepackage[dvipsnames]{xcolor}
\usepackage{soul}
\usepackage{enumitem}
\usepackage{lipsum}

\usepackage[normalem]{ulem}

\usepackage[colorlinks=true,
            linkcolor=blue,
            citecolor=blue,
            urlcolor=blue]{hyperref}

\usepackage{lineno}

\newcommand{\addComment}[2]{
  \expandafter\newcommand\csname #1\endcsname[1]{{\bf \color{#2} \capitalisewords{#1}:\,##1}}
  \expandafter\newcommand\csname #1cor\endcsname[2]{{\color{#2} \capitalisewords{#1}:\,\st{##1}{\bf ##2}}}
  \expandafter\newcommand\csname #1color\endcsname{#2}
}

\addComment{cole}{red}
\addComment{cris}{blue} 
\addComment{ale}{ForestGreen} 
\addComment{marco}{red}
\addComment{simone}{cyan}

\begin{document}

\preprint{APS/123-QED}

\title{Symbolic Extraction of Non-Perturbative\\
Transverse-Momentum-Dependent Distributions from Drell-Yan Data}

\author{
Cole Granger$^{1}$,
Alessandro Bacchetta$^{2,3}$,
Valerio Bertone$^{4}$,
Chiara Bissolotti$^{5}$,
Matteo Cerutti$^{4}$,
Marco Radici$^{3}$,
Simone Rodini$^{2,3}$,
Lorenzo Rossi$^{6,7}$,
Cristiano Fanelli$^{1}$
}

\email{cfanelli@wm.edu}

\affiliation{
$^{1}$ William \& Mary, School of Computing, Data Sciences and Physics, Williamsburg, Virginia, 23185 \\
$^{2}$ Dipartimento di Fisica ``A. Volta," Universit\`a di Pavia, via Bassi 6, I-27100 Pavia, Italy \\
$^{3}$ INFN, Sezione di Pavia, via Bassi 6, I-27100 Pavia, Italy \\
$^{4}$ IRFU, CEA, Universit\'e Paris-Saclay, F-91191 Gif-sur-Yvette, France \\
$^{5}$ Argonne National Laboratory, PHY Division, Lemont, IL, USA \\
$^{6}$ Dipartimento di Fisica, Universit\`a di Milano, Via Celoria 16, I-20133 Milan, Italy \\
$^{7}$ INFN, Sezione di Milano, Via Celoria 16, I-20133 Milan, Italy
}

\date{\today}


\begin{abstract}
We present an analytical parametrization of the non-perturbative transverse-momentum-dependent (TMD) parton distribution function of unpolarized quarks, extracted from Drell–Yan data using a combination of neural-network fitting and symbolic regression. A factorized neural network is trained directly against experimental cross-section data from fixed-target, Tevatron, RHIC, and LHC experiments at next-to-next-to-next-to-leading logarithmic accuracy, and symbolic regression is subsequently applied to each network component to discover compact analytical expressions. The final formula is selected from a Pareto front in the space of expression complexity and experimental $\chi^2$, yielding a closed-form non-perturbative function with 9 free numerical constants that achieves $\chi^2/\text{ndf} = 1.040$ over 482 data points.
A non-trivial $x$--$b_T$ cross term is retained even under a sparsity prior that biases it toward zero, indicating a mild but genuine correlation between the longitudinal momentum fraction and the transverse momentum.
This work demonstrates that symbolic regression is a viable tool for bridging flexible machine-learning fits and interpretable analytical TMD parametrizations,
and opens a systematic path toward data-driven discovery of specific features of non-perturbative QCD. 
\end{abstract}

\maketitle


\input{1_introduction}

\input{2_theoretical_framework}

\input{3_experimental_data}

\input{4_methodology}

\input{5_results}

\input{6_discussion}

\input{7_conclusion}

\input{acknowledgments}

\bibliographystyle{apsrev4-2}
\bibliography{apssamp}

\newpage

\end{document}

%% file: 1_introduction.tex
\section{\label{sec:intro}Introduction}

Transverse-momentum-dependent (TMD) distributions provide a three-dimensional momentum-space description of hadrons by encoding correlations between the intrinsic transverse momentum of partons, their longitudinal momentum fraction, the hard scale of the process, and the involved spins, although we will focus here only on unpolarized TMDs (see, e.g., \cite{Collins1985, Collins2011, Angeles_Martinez_2015,Boussarie:2023izj}).
They consist of a perturbative and a nonperturbative component ($f_{NP}$), where the former can be computed within perturbative QCD, while the latter encapsulates long-distance physics and must be inferred from experimental data, typically through fits to processes such as Drell–Yan (DY) production. Conventional parameterizations, often based on Gaussian or exponential functional forms supplemented by relatively low-dimensional kinematic dependence (see, e.g., ~\cite{Bacchetta:2017gcc,Scimemi:2017etj,Bertone:2019nxa,Scimemi:2019cmh,Bacchetta:2019sam,Bury:2022czx,Bacchetta:2022awv,Moos:2023yfa,Bacchetta:2024qr, DYTurbo_fit}), provide a practical and widely used framework, but may introduce functional assumptions that can influence the flexibility of the extraction. More recently, the MAP Collaboration developed a neural-network approach~\cite{Bacchetta:2025ara} to reduce such model dependence. 
However, the more flexible parametrization is obtained
at the cost of a higher-dimensional representation that can be more challenging to interpret or directly compare with traditional analyses.




Symbolic regression has recently emerged as a useful tool for constructing interpretable analytic forms in hadron and collider phenomenology. Recent applications include symbolic parametrizations of generalized parton distributions from lattice-QCD and phenomenological inputs, with emphasis on extrapolation behavior, factorization tests, and symbolic-convergence diagnostics~\cite{Dotson:2025omi}; compact analytic approximations for precision LHC phenomenology, validated through the rediscovery of known QED angular distributions and then applied to Drell--Yan structure functions~\cite{Morales-Alvarado:2024jrk, Bendavid:2025urn}; and fragmentation-function-inspired forms inferred from semi-inclusive DIS multiplicities, where the learned expressions resemble Lund-type fragmentation models~\cite{Makke:2025zoy}. 
These studies demonstrate the potential of symbolic regression to expose simple structures behind numerical, simulated, or data-driven inputs. Other work has applied an Agentic workflow utilizing OpenAI Codex to suggest formulations for the nonperturbative ansatze \cite{Kang:2026mod}.
In the TMD case, however, the relevant nonperturbative functions are not directly observed quantities: they enter physical cross sections through TMD evolution, flavor sums, and kinematic convolutions, and must ultimately be assessed against correlated experimental datasets.

This makes it essential to assess symbolic candidates not only as functions, but through their impact on the measured cross sections. The present work follows this observable-level strategy.
In this work, we introduce symbolic regression as a tool for TMD nonperturbative function extraction and report the first analytical  formula for $f_{\rm NP}$ obtained directly from data rather than postulated, within a factorized enveloping ansatz motivated  by symbolic-regression tractability (Sec.~\ref{sec:method}). We train a factorized neural network on a global Drell–Yan dataset 
evaluated at N$^3$LL accuracy within the NangaParbat 
framework~\cite{Bacchetta:2022awv,NangaParbat} and apply symbolic regression to extract compact analytical expressions, selecting candidates from a combined complexity--$\chi^2$ Pareto front validated against  experimental data. 
The resulting expression --- a linear $x$ term, a linear $b_T$ term corrected by a factor, and a small genuinely non-separable cross-term --- achieves $\chi^2/\text{ndf} = 1.040 \pm 0.065$ over 482 data points.
The formula is compact enough to be re-implemented directly in established  global-fit codes and reveals interpretable structure.

%% file: 2_theoretical_framework.tex
\section{Theoretical Framework}{}

We focus on the DY process $h_A(P_A) + h_B(P_B) \to \ell^+ \ell^- + X$,
where $P_A$ and $P_B$ denote the four-momenta of the incoming hadrons
$h_A$ and $h_B$, whose masses are $M_A$ and $M_B$, respectively. The
hadronic center-of-mass energy is given by $s = (P_A + P_B)^2$, while the
lepton pair has total four-momentum $q$ and invariant mass $Q^2 = q^2$,
with $Q \gg M_{A,B}$. In the region where the transverse momentum of the
lepton pair, measured with respect to the collision axis, satisfies
$|\mathbf{q}_T| \equiv q_T \ll Q$, the cross section admits a
TMD-factorised representation. In this limit, the factorised expression for
the differential cross section is
%
%
\begin{equation}
\label{eq:DY_xsec}
\begin{aligned}
&\frac{d\sigma^{\mathrm{DY}}}{dq_T\, dy\, dQ}
= \frac{8\pi\alpha^2 q_T}{9Q^3}\,
  \mathcal{P}\, x_A x_B\, H^{\mathrm{DY}}(Q, \mu)
  \sum_a c_a(Q^2)  \\
& \times\! \int_0^\infty\! db_T\, b_T\,
  J_0\!\left(b_T q_T\right)\!
  \hat{f}_1^a\!\left(x_A, b_T^2;\, \mu, \zeta_A\right)\!
  \hat{f}_1^{\bar{a}}\!\left(x_B, b_T^2;\, \mu, \zeta_B\right),
\end{aligned}
\end{equation}
where, $y = \ln\sqrt{(q^0+q^z)/(q^0-q^z)}$ denotes the rapidity of the lepton
pair, while $\alpha$ is the electromagnetic coupling. The factor
$\mathcal{P}$ accounts for the reduction of the available phase space due to
possible cuts on the final-state leptons. The variables
$x_{A,B} = Q e^{\pm y}/\sqrt{s}$ represent the longitudinal momentum
fractions entering the two distributions. The function $H^{\mathrm{DY}}$ is the
hard factor, which contains the perturbative virtual corrections to the
short-distance scattering. Finally, the sum extends over all active quark
flavours $a$, and the coefficients $c_a$ collect the corresponding
electroweak charge factors~\cite{Bacchetta:2019sam}.

The quantity $\hat{f}_1^a$ appearing in Eq.~\eqref{eq:DY_xsec} is the Fourier transform of
the unpolarised TMD Parton Distribution Function (PDF) of quark flavour $a$. It depends on the quark
longitudinal momentum fraction $x$ and on the variable
$b_T = |\mathbf{b}_T|$, where $\mathbf{b}_T$ is the Fourier conjugate to the intrinsic transverse momentum $\mathbf{k}_\perp$ of the quark. In addition, it depends on
the renormalisation scale $\mu$ and on the rapidity scale $\zeta$, with the latter satisfying the constraint $\zeta_A \zeta_B = Q^4$ . These scale dependences arise
from the subtraction of ultraviolet and rapidity divergences \cite{Collins2011} and are
governed by the corresponding evolution equations. Omitting variables
and indices that are not relevant for the following discussion, these
equations read
\begin{align}
\label{eq:evolution}
\frac{\partial \ln \hat{f}_1}{\partial \ln\mu} &= \gamma(\mu, \zeta), \quad
\frac{\partial \ln \hat{f}_1}{\partial \ln\sqrt{\zeta}} = K(\mu), \\
\frac{\partial K}{\partial \ln\mu}
&= \frac{\partial \gamma}{\partial \ln\sqrt{\zeta}}
= -\gamma_K(\alpha_s(\mu)),
\end{align}
where $\gamma$ and $K$ are the anomalous dimensions of the
renormalisation group and Collins--Soper equations, respectively,
and $\gamma_K$ is the so-called cusp anomalous dimension which relates
the cross-derivatives of $\hat{f}_1$.

Starting from a given pair of initial scales $(\mu_i,\zeta_i)$, the solution
of the evolution equations determines the TMD PDF at any final scales
$(\mu_f,\zeta_f)$. At small transverse separations $b_T$, the TMD PDF
$\hat{f}_1$ can also be matched onto the corresponding unpolarised collinear
PDF $f_1$ through an operator product expansion involving perturbatively
calculable matching coefficients $C$.

The resulting TMD PDF, evolved to the final scales $(\mu_f, \zeta_f)$, can
therefore be written as
\begin{align}
\label{eq:TMD_PDF}
&\hat{f}_1(x, b_T;\, \mu_f, \zeta_f)
= \left(C \otimes f_1\right)\!(x, b_T;\, \mu_i, \zeta_i) \nonumber \\
& \quad  \times \exp\!\Bigg{\{}
    K(\mu_i)\ln\frac{\sqrt{\zeta_f}}{\sqrt{\zeta_i}} \nonumber \\
& \qquad +\! \int_{\mu_i}^{\mu_f} \!\frac{d\mu}{\mu}\,
        \gamma_F(\alpha_s(\mu)) - \!\int_{\mu_i}^{\mu_f}\! \frac{d\mu}{\mu}\gamma_K(\alpha_s(\mu))\ln\frac{\sqrt{\zeta_f}}{\mu}
  \Bigg{\}},
\end{align}
where $\gamma_F(\alpha_s(\mu)) = \gamma(\mu, \mu^2)$ and $\otimes$
indicates the Mellin convolution over the longitudinal momentum fraction
$x$. A natural choice for the initial scales is
$\mu_i = \sqrt{\zeta_i} \equiv \mu_b = 2e^{-\gamma_E}/b_T$, where
$\gamma_E$ is the Euler constant. With this choice, large logarithms are
avoided both in the Collins--Soper kernel $K$ and in the matching
coefficients $C$. The TMD PDF in Eq.~\eqref{eq:TMD_PDF} includes the
resummation of large logarithms of $b_T$ to all orders in perturbation
theory. A given logarithmic accuracy implies that each ingredient in
Eq.~\eqref{eq:TMD_PDF} must be computed to the appropriate perturbative
accuracy. The present extraction incorporates all the necessary
ingredients to reach N$^3$LL accuracy.

The introduction of the scale $\mu_b \sim 1/b_T$ requires a
prescription to avoid integrating in Eq.~\eqref{eq:DY_xsec} over the
QCD Landau pole ($\Lambda_{\mathrm{QCD}}$) in the large-$b_T$ region.
To this purpose, we adopt the same choice as in Refs.~\cite{Bacchetta:2017gcc,Bacchetta:2019sam,Bacchetta:2022awv,Bacchetta:2024qr,Bacchetta:2025ara,Cerutti:2022lmb,Rossi:2025pwh,Bacchetta:2024yzl,Cerutti:2026apy,Anedda:2026cox} and we replace $\mu_b$ with
$\mu_{b^*} = 2e^{-\gamma_E}/b^*$, where
\begin{equation}
\label{eq:bstar}
b^*(b_T, b_{\min}, b_{\max})
= b_{\max}
  \left(
    \frac{1 - e^{-b_T^4/b_{\max}^4}}
         {1 - e^{-b_T^4/b_{\min}^4}}
  \right)^{1/4},
\end{equation}
with
\begin{equation}
\label{eq:bmin_bmax}
b_{\max} =2e^{-\gamma_E}\, \text{GeV}^{-1},
\qquad
b_{\min} = \frac{2e^{-\gamma_E}}{\mu_f}.
\end{equation}
This prescription guarantees that the variable $b^*$ rapidly saturates to
$b_{\max}$ at large values of $b_T$, thereby preventing $\mu_{b^*}$ from
reaching $\Lambda_{\mathrm{QCD}}$. At the same time, the use of $b^*$ induces
spurious power corrections of the form $(\Lambda_{\mathrm{QCD}}/q_T)^k$ \cite{Kuleza_2002, Kuleza_2004, Catani_1996, Laenen_2000},
with $k > 0$. In the region $q_T \simeq \Lambda_{\mathrm{QCD}}$, these
power corrections are no longer negligible and can be modelled by including in
Eq.~\eqref{eq:TMD_PDF} the nonperturbative function $f_{\mathrm{NP}}$ as
follows:
\begin{align}
\label{eq:TMD_NP}
&\hat{f}_1(x, b_T;\, \mu_f, \zeta_f)
= \left(C \otimes f_1\right)\!(x, b_T;\, \mu_{b^*}, \mu_{b^*}^2)
\nonumber \\
&\quad \times \exp\!\Bigg{\{}
    K(b^*, \mu_{b^*})\ln\frac{\sqrt{\zeta_f}}{\mu_{b^*}}
\nonumber \\
&\qquad +\! \int_{\mu_{b^*}}^{\mu_f} \!\frac{d\mu}{\mu}\,
      \gamma_F(\alpha_s(\mu)) - \!\int_{\mu_{b^*}}^{\mu_f} \!\frac{d\mu}{\mu}\,\gamma_K(\alpha_s(\mu))\ln\frac{\sqrt{\zeta_f}}{\mu}
  \Bigg{\}}
\nonumber \\
&\qquad \quad \times f_{\mathrm{NP}}(x, b_T;\, \zeta_f).
\end{align}

The nonperturbative function must satisfy the condition
$f_{\mathrm{NP}} \to 1$ for $b_T \to 0$ in order to recover the
perturbative regime. It must also guarantee that the TMD PDF is
suppressed for large values of $b_T$ and $\zeta_f$.  

%% file: 3_experimental_data.tex
\section{Experimental Data}\label{sec:data}
The analysis uses the same experimental dataset used in  \cite{Bacchetta:2025ara}, comprising Drell–Yan transverse-momentum measurements from fixed-target experiments at Fermilab (E288 \cite{E228}, E605 \cite{E605}, E772 \cite{E772}), from the Tevatron (CDF \cite{CDF1, CDF2}, D0 \cite{D01, D02, D03}) and RHIC colliders (STAR \cite{STAR}), and from the LHC (LHCb \cite{LHCb1, LHCb2, LHCb3}, CMS \cite{CMS1, CMS2, CMS3}, ATLAS \cite{ATLAS1, ATLAS2, ATLAS3}). 
A summary of the datasets included from the various experiments is provided in Table~\ref{tab:experimental_data}.
The full dataset consists of 482 data points from 9 experiments. Each experiment is partitioned into kinematic bins; bins belonging to different experiments are treated as distinct even when they cover similar kinematic ranges. In total, this defines 59 subsets of experiment-specific data.

The following kinematic selections are applied uniformly:
(i) a cut in the region $q_T / Q < 0.2$, ensuring that the observed transverse momentum is genuinely in the TMD regime; (ii) a veto of the $\Upsilon$-resonance region, excluding data with $9\ \text{GeV} < Q < 11\ \text{GeV}$.
\begin{table}[h]
\centering
\caption{Dataset summary across experiments. Dataset follows \cite{Bacchetta:2025ara}.}

\begin{tabular}{llc}
\hline
\textbf{Source} & \textbf{Experiment} & $\boldsymbol{N_{\rm dat}}$ \\
\hline
Fermilab & E288 & 130 \\
 & E605 & 50 \\
 & E772 & 53 \\
Tevatron & CDF Run I \& II & 51 \\
 & D0 Run I \& II & 20 \\
RHIC & STAR & 7 \\
LHC & ATLAS 7, 8, 13 TeV & 72 \\
 & CMS 7, 8, 13 TeV & 78 \\
 & LHCb 7, 8, 13 TeV & 21 \\
\hline
& \textbf{Total} & \textbf{482} \\
\hline
\end{tabular}
\label{tab:experimental_data}
\end{table}

 The experimental covariance matrices, including all sources of correlated systematic uncertainty, are provided by the convolution skeleton described in Sec.~\ref{ss:experim_data_conv_skeleton}, where the treatment of correlated systematics is detailed.

%% file: 4_methodology.tex
\section{Methodology}\label{sec:method}
\subsection{Parametric Form of the Non-Perturbative TMD Parton Distribution Function}
\label{sec:method:paramterization}
The non-perturbative component of the unpolarized TMD PDF is parametrized as
\begin{equation}
f_{\rm NP}(x, b_T; \zeta) = \exp\Bigl[-g_2^2(b_T)\, b_T^2 \ln\bigl(\zeta / Q_0^2\bigr)\Bigr] \cdot R(x, b_T),
\end{equation}
where the exponential describes the nonperturbative part of the Collins-Soper kernel through the parametric function $g_2(b_T)$, with
$Q_0^2 = 1\ \text{GeV}^2$ a fixed reference scale. The function $R(x,b_T)$ encodes the intrinsic transverse-momentum shape at the initial scale and is normalized so that $R(x, 0) = 1$ for all $x$. 

Its full form is written as 
\begin{equation}
R(x, b_T) = \exp\bigl[h_x(x) \, h_b(b_T)  + h_{xb}(x, b_T)\bigr], 
\label{eq:parametrize}
\end{equation}
where the cross term $h_{xb}(x,b_T)$, which captures $x$--$b_T$ correlations, is included throughout the training. The factorization in Eq.~(\ref{eq:parametrize}) is motivated both by physics and by symbolic-regression tractability. Symbolic regression scales poorly with input dimensionality, and fitting $R(x,b_T)$ directly as a single two-dimensional network produces expressions too complex for interpretable distillation. In separate attempts to fit the neural network output directly, no candidate expression achieved $\chi^2/N_\mathrm{dat} < 1.3$ on the experimental data. Decomposing $R$ into two one-dimensional networks $h_x(x)$, $h_b(b_T)$ and a small cross term $h_{xb}(x,b_T)$ reduces the symbolic search to a tractable problem 
while preserving the expressive power needed to match the data. 
To prevent $h_{xb}$ from absorbing residual signal at the expense of the one-dimensional components, three stabilizers are imposed jointly: (i)~$L_1$ regularization is applied to the $h_{xb}$ network parameters (Cranmer sparsity prior~\cite{cranmer2020discoveringsymbolicmodelsdeep}), biasing the cross term toward zero whenever the data do not require it; (ii)~$h_{xb}$ is given limited capacity (one hidden layer of four neurons), imposing a structural cap on the amount of residual signal it can absorb; and~(iii)~the experimental $\chi^2$ is computed with normalization applied to experiment-specific data subsets such that no single dataset can pull $h_{xb}$ disproportionately (see Sec.~\ref{sec:method:training}). Under these stabilizers, the joint optimization recovers a small non-zero $h_{xb}$ whose magnitude is data-determined rather than initialization-determined.

The architectural boundary conditions $h_b(0) = 0$ and $h_{xb}(x,0) = 0$ are enforced by construction in the neural-network parametrization (see Sec.~\ref{sec:method:nn}), so that $R(x, b_T=0) = 1$ holds exactly throughout the training. The exponential form of $R$ yields a positive function on the entire $(x, b_T)$ domain.
\subsection{Neural Network (NN)}\label{sec:method:nn}
Each component of the parametrization in Eq.~\eqref{eq:parametrize} is represented by a shallow multilayer perceptron (MLP) with $\tanh$ activation:
\begin{enumerate}[label=(\roman*),noitemsep,topsep=0pt,parsep=0pt,partopsep=0pt]
    \item $h_x(x)$: single hidden layer, $N_h = 8$ neurons, input $x \in (0,1)$.
    \item $h_b(b_T)$: single hidden layer, $N_h = 8$ neurons. The boundary condition $h_b(0) = 0$ is enforced architecturally by subtracting the network's output at $b_T = 0$: $h_b(b_T) \equiv \hat{h}_b(b_T) - \hat{h}_b(0)$.
    \item $g_2(b_T)$: single hidden layer, $N_h = 4$ neurons. This allows $g_2$ to develop mild $b_T$-dependence beyond a constant while remaining $x$-independent, consistent with the leading Collins--Soper non-perturbative kernel. The network output is passed through a softplus activation to enforce positivity of the evolution coupling on the trained kinematic domain.
    \item $h_{xb}(x, b_T)$: single hidden layer, $N_h = 4$ neurons. The cross term is trained jointly with the other components from epoch zero, but two structural constraints prevent it from absorbing signal that should belong to the factorized backbone: an $L_1$ penalty on its weights (with strength $\lambda_{L_1} = 10^{-4}$) biases it toward zero, and the limited hidden width caps its expressive capacity. The architectural boundary condition $h_{xb}(x, 0) = 0$ is enforced by output subtraction $h_{xb}(x, b_T) \equiv \hat{h}_{xb}(x, b_T) - \hat{h}_{xb}(x, 0)$, mirroring the $h_b$ construction. This guarantees that the matching condition $R(x, 0) = 1$ holds exactly throughout training.
\end{enumerate}
Hard physical boundary conditions ($R = 0$ for $b_T > 10$ and for $x \geq 1$) are applied as masks during evaluation, ensuring the symbolic formula is never extrapolated into unwanted or unphysical kinematic regions.

\subsection{Experimental Data and the Convolution Skeleton}
\label{ss:experim_data_conv_skeleton}
Theory predictions for Drell–Yan transverse-momentum distributions are obtained via Ogata quadrature in $b_T$ space, where $b_T$ is Fourier conjugate to the observed transverse momentum $q_T$ \cite{ogata2005numerical}. 
To make gradient-based optimization tractable, the convolution integrals are precomputed and stored as a convolution skeleton, namely an HDF5 file containing, for each contribution to each data bin, the quadrature weights, phase-space factors, momentum fractions $(x_1,x_2)$, values of $b_T$, and the evolution scale $\zeta$. At each training step, $f_{\rm NP}$ is evaluated at these precomputed $(x, b_T, \zeta)$ points, multiplied by the stored weights, and summed to assemble theory predictions without recomputing any quadrature grid.

Each data subset carries its own correlated systematic uncertainties, between three and six sources per data point depending on the experiment; these enter the block of the covariance matrix belonging to that subset and are not shared across experiments. They are fully contained in the experimental covariance matrix ${\cal C}$ and are treated exactly through the analytic nuisance-parameter profiling described below. 

The matrix ${\cal C}$, which includes all correlated systematic uncertainties, is decomposed via Cholesky factorization
${\cal C} = L L^{\mathsf{T}}$, giving the figure of merit
\begin{equation}
  \chi^2 = \left\lVert L^{-1}\left(\mathbf{t} - \mathbf{d}\right)\right\rVert_2^2 ,
  \label{eq:chi2_cholesky}
\end{equation}
where $L$ is the lower Cholesky factor of the full covariance and $\mathbf{t}$, $\mathbf{d}$ are the vectors of theory predictions and experimental measurements. This figure of merit, identical to that of the reference fit~\cite{Bacchetta:2025ara}, is the goodness of fit we report; the training loss is built from it as described in~\ref{sec:method:training} but supplements it with regularization terms that are not included in any reported figure of merit. The correlated systematic uncertainties are fully contained in ${\cal C}$; we evaluate this $\chi^2$ through the equivalent analytic nuisance-parameter profiling, following the same treatment as the reference fit~\cite{Bacchetta:2025ara}, and report the fitted nuisance parameters post-fit for the correlated/uncorrelated $\chi^2$ decomposition and the pulls.
(Sec.~\ref{sec:results}).

\subsection{Training}\label{sec:method:training}
All training is done using the PyTorch~\cite{pytorch} library. The factorized model is trained directly against the experimental $\chi^2$. We explored initialization strategies including warm-starting from a fully unconstrained two-dimensional NN, but found that the gradient structure learned by the reference fit could not be effectively transferred to the constrained factorized form: distillation attempts yielded poor convergence and suboptimal $\chi^2$. This incompatibility arises because the factorized exponential parameterization in Eq.~\eqref{eq:parametrize} imposes a fundamentally different loss landscape than an unconstrained 2D network. Consequently, the factorized model finds its own solution directly on the experimental loss landscape from the outset.
The total loss at each step is
\begin{equation}
\mathcal{L} \;=\; \mathcal{L}^{\rm bal}_{\chi^2} \;+\; \lambda_{L_1}\,\|\theta_{h_{xb}}\|_1 \;+\; \lambda_{L_2}\,\|\theta_{\rm shape}\|^2.
\label{eq:training-loss}
\end{equation}

The three terms in Eq.~(\ref{eq:training-loss}) correspond to distinct objectives.
The first, $\mathcal{L}^{\rm bal}_{\chi^2}$ is given by
\begin{equation}
\mathcal{L}^{\rm bal}_{\chi^2} = \frac{1}{N_{\rm exp}}\sum_{s} w_s \frac{\chi^2_s}{N_s}
  + \lambda_{\rm bal}\,\frac{1}{N_{\rm exp}}\sum_{s}
    \left(\frac{\chi^2_s}{N_s} - \langle{\chi^2/N}\rangle\right)^2,
\end{equation}
where $s$ runs over the $N_{\rm exp} = 59$ experiment-specific data subsets, $N_s$ is the number of data points in subset $s$, $\langle\chi^2/N\rangle = N_{\rm exp}^{-1}\sum_s \chi_s^2/N_s$ is the mean reduced $\chi^2$ over the subsets, and $w_s$ are the subset weights defined below.

The $\mathcal{L}^{\rm bal}_{\chi^2}$ is a $\chi^2$ loss balanced across
data subsets: each subset's $\chi^2_s$ is normalized by its number of data points $N_s$, and the penalty term with $\lambda_{\rm bal} = 5.0$ drives the reduced $\chi^2_s$ values of the subsets toward their cross-experiment mean, preventing the optimizer from sacrificing poorly fitted low-statistics experiments to minimize the global loss. The second term, $\lambda_{L_1}\|\theta_{h_{xb}}\|_1$ with $\lambda_{L_1} = 10^{-4}$, is an $L_1$ sparsity penalty applied exclusively to the $h_{xb}$ network parameters, biasing the cross term toward zero unless the data require it. The third term, $\lambda_{L_2}\|\theta_{\rm shape}\|^2$ with $\lambda_{L_2} = 10^{-4}$, is a mild $L_2$ weight decay  
on the parameters of the remaining shape networks ($h_x$, $h_b$, $g_2$), discouraging arbitrarily large weights without strongly constraining the fit. To address residual tension in fixed-target low-$Q$ Drell--Yan data,  the weights are set to $w_s = 3$ for the E772 subsets at $Q \in [5,9]$~GeV and $w_s = 1$ for all other subsets.

Separate learning rates are assigned to $g_2$ ($\eta = 10^{-4}$) and to the shape networks $h_x, h_b, h_{xb}$ ($\eta = 2 \times 10^{-4}$). The learning rate is halved whenever the training loss plateaus for 800 consecutive epochs. The full dataset is used for training with no held-out validation split. Training runs for up to $100{,}000$ epochs with gradient clipping at norm~$1.0$, and all computations are performed in double precision on an NVIDIA~4090 GPU.

We do not hold out a random subset of bins during training. Instead, three structural choices protect the extraction against overfitting. First, the NN is intentionally kept small --- each component ($h_x$, $h_b$, $g_2$, $h_{xb}$) is a single-hidden-layer MLP with only $4$ -- $8$ neurons (see Sect.~\ref{sec:method:nn}), totaling $\mathcal{O}(100)$ parameters --- so the architecture itself has limited capacity to memorize noise. Second, the symbolic compression itself reduces $\mathcal{O}(100)$ network parameters to nine analytic constants while degrading the global $\chi^2/\text{ndf}$ by only $\sim 8\%$ --- a compression ratio incompatible with a network that has fit noise rather than the underlying non-perturbative structure.
Third, the Pareto-knee analysis of Sec.~\ref{sec:method:pareto} act as another form of regularization of the symbolic solution. 

The final checkpoint is selected by a minimax criterion on the reduced $\chi^2$ of each data subset: the procedure is stopped at the epoch at which the worst-performing data subset achieves its lowest reduced $\chi^2$, where the identity of the worst subset is re-evaluated at each epoch as the fit evolves. Formally, denoting by $\chi^2_{s,\tau} /N_{s}$ a proxy for the reduced $\chi^2$ of subset~$s$ at epoch~$\tau$, with $N_{s}$ being the number of data points for a data subset $s$ of a given experiment, the selected epoch $\tau^\star$ satisfies
\begin{equation}
\tau^\star \;=\; \arg\min_\tau \max_s \,\chi^2_{s,\tau}/N_{s}.
\label{eq:minimax-selection}
\end{equation}
This criterion guards against checkpoints that achieve a low mean $\chi^2$ by fitting the bulk of experiments well while leaving one or two experiments poorly described, a failure mode that is mitigated by normalizing the loss by subset, but not eliminated.

\subsection{Symbolic Regression}
\label{ss:symbolic_regression}
After training, each component ($h_x$, $h_b$, $g_2$) is evaluated on a dense grid and submitted independently to PySR \cite{cranmer2023interpretable}, a Julia-based evolutionary algorithm for symbolic regression. PySR evolves a population of candidate analytical expressions via genetic-programming operations, crossover, mutation, and algebraic simplification, and returns a Pareto front of equations trading expression complexity (node count in the expression tree) against mean-squared error on the evaluation grid.

The symbolic regression, and the combined-$\chi^2$ selection that follows, are restricted to the fitted range $b_T < b_T^{\rm cut} = 2.5~\mathrm{GeV}^{-1}$. This cutoff --- distinct from the $b^*$-prescription parameter $b_{\max}=2e^{-\gamma_E}~\mathrm{GeV}^{-1}$ of Eq.~(6) --- is set by the reach of the Drell--Yan data: beyond $b_T\simeq2.5~\mathrm{GeV}^{-1}$ the Collins--Soper kernel non-perturbative part $\exp[-g_2^2 b_T^2\ln(\zeta/Q_0^2)]$ suppresses $f_{\rm NP}$ to the point that the data carry essentially no constraint, and unconstrained large-$b_T$ behavior would only inflate expression complexity. Fitting within $b_T<b_T^{\rm cut}$ keeps the extracted formulas faithful to the data-constrained region.

\paragraph{Normalization convention.}
Because $h_x(x) \, h_b(b_T)$ is invariant under the rescaling $(h_x \, h_b) \to (c \, h_x \, h_b/c)$, before symbolic regression, a canonical normalization is imposed on the trained network outputs to break the multiplicative degeneracy of the product $h_x \, h_b$: the rescaling constant $\alpha = -1/h_x(x_{\rm ref})$ is computed at $x_{\rm ref} = 0.1$, and the components are rescaled as $h_x \to \alpha \, h_x$, $h_{b} \to h_b/\alpha$.

The value of $\alpha$ is recorded alongside the extracted formulas to allow reconstruction of the original (unnormalized) network. The rescaling constant is fixed by the convention $h_x(0.1) = -1$, where $x_{\rm ref} = 0.1$ is chosen as a point well within the data-constrained region where $h_x$ is robustly non-zero, and the negative sign reflects the physical requirement that the exponent $h_x \, h_b$ be negative for $b_T > 0$, ensuring $R(x,b_T) < 1$ at large $b_T$.

\paragraph{Search configuration.}
The search uses 40 parallel subpopulations of 150 candidate expressions, run for 500 generations with a maximum tree size of 25–35 nodes. Binary operators are ${+, -, \times, \div}$ and unary operators are ${\log, \sqrt{\phantom{x}}, \exp, (\cdot)^2, |\cdot|}$. Nested transcendental compositions (\textit{e.g.},\ $\exp(\exp(\cdot))$, $\log(\log(\cdot))$) are forbidden via nested-operator constraints. A parsimony coefficient of $5 \times 10^{-4}$ --- $8 \times 10^{-4}$ penalizes complexity during the search. Physics-motivated sampling weights emphasize $x  < 0.5$  and $b_T < 2.5$ regions, where experimental constraints are strongest.

\subsection{Pareto Selection}\label{sec:method:pareto}
The final analytical formula is selected in two stages:

\paragraph{Component-wise symbolic regression}
PySR is run independently on each component ($h_x$, $h_b$, $h_{xb}$, $g_2$), fitting candidate expressions against the trained NN outputs evaluated on a dense grid using mean-squared error (MSE). Each run produces a Pareto front of equations in the (complexity, MSE) plane, from which a subset of candidates is retained for the next stage.

\paragraph{Combined experimental $\chi^2$ Pareto}
All combinations of Pareto-front candidates across the four
components are assembled into a complete $f_{\rm NP}$ and evaluated against the experimental data through the full convolution skeleton. For a combination indexed by $(i, j, k, \ell)$, the non-perturbative function is
%
%
\begin{equation}
\begin{aligned}
&f_{\rm NP}(x_1,x_2,b,\zeta;i,j,k,\ell)
= \\
&\quad \exp\Biggl[
\bigl(h_x^{(i)}(x_1)+h_x^{(i)}(x_2)\bigr)h_b^{(j)}(b_T) \\
&\qquad \quad + h_{xb}^{(\ell)}(x_1, b_T)
+ h_{xb}^{(\ell)}(x_2, b_T) \\
&\qquad \quad -\bigl(g_2^{(k)}(b)\bigr)^2 b_T^2
\ln\!\left(\frac{\zeta}{Q_0^2}\right)
\Biggr],
\end{aligned}
\end{equation}
where the same $h_x^{(i)}$ is applied to both incoming momentum fractions $x_1$ and $x_2$ that describe the Drell-Yan process. Theory predictions are assembled via the precomputed skeleton weights and compared to the experimental data to yield a $\chi^2/\text{ndf}$ for each combination, recorded alongside the total expression complexity (sum of node counts across the four components).

The Pareto analysis is considered over ``\textit{minimax subset $\chi^2/N_s$ at fixed complexity}’’ which selects the candidate that minimizes the largest data subset $\chi^2/N_{s}$ across all~$59$ kinematic bins from across 9 different experiments (cf. Table \ref{tab:experimental_data}). We adopt the worst-data-subset criterion for the final formula because the balanced-loss training (see Sec.~\ref{sec:method:training}) already targets uniform subset fit quality; the natural complement at the selection stage is therefore to enforce uniformity at the formula level as well. 

This is intended to avoid solutions for which the global reduced $\chi^2$ is statistically compatible with unity, while individual data subsets still exhibit large $\chi^2/N_{s}$. Such cases can occur when localized failures in specific kinematic bins are masked by good average performance over the full dataset. The worst-data-subset selection criterion directly penalizes these localized discrepancies and therefore provides a more stringent robustness test.

%% file: 5_results.tex
\section{Results}\label{sec:results}

\begin{figure}[!htbp]
    \centering
    \includegraphics[width=0.475\textwidth]{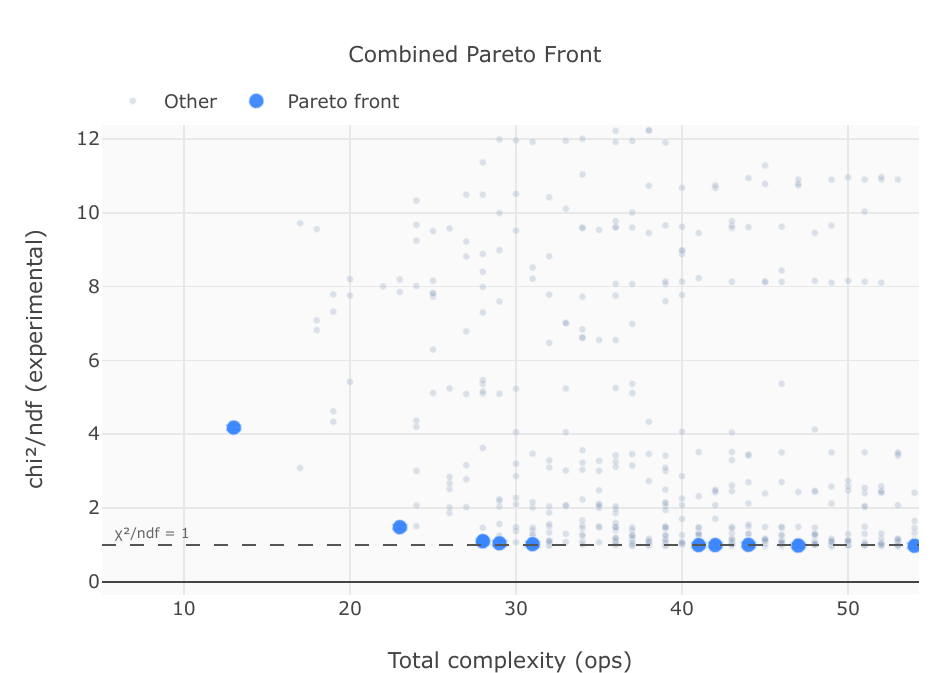}
    \caption{The Pareto front: candidate symbolic parametrizations are compared in terms of total analytical complexity, measured in operations, and global $\chi^2/\mathrm{ndf}$. Gray points denote suboptimal solutions, while blue points denote Pareto-optimal solutions. The horizontal dashed line marks $\chi^2/\mathrm{ndf}=1$, providing a reference for statistically acceptable agreement with the data.
    }
    \label{fig:pareto}
\end{figure}

The Pareto-optimal solutions define the set of parametrizations for which no alternative achieves both lower analytical complexity and better agreement with the data. An interactive Pareto explorer is available on GitHub~\cite{pareto_explorer}. Figure~\ref{fig:pareto} shows a zoomed-in view of the Pareto front included in the interactive explorer. 

This representation makes explicit the trade-off between fit quality and interpretability: while increasing the complexity generally improves the description of the data, the improvement becomes marginal beyond moderate complexity. We therefore select a solution on the Pareto front that provides a statistically acceptable description of the data while avoiding unnecessary analytical complexity.

\begin{figure*}[htbp]
    \centering
    \includegraphics[width=0.85\textwidth]{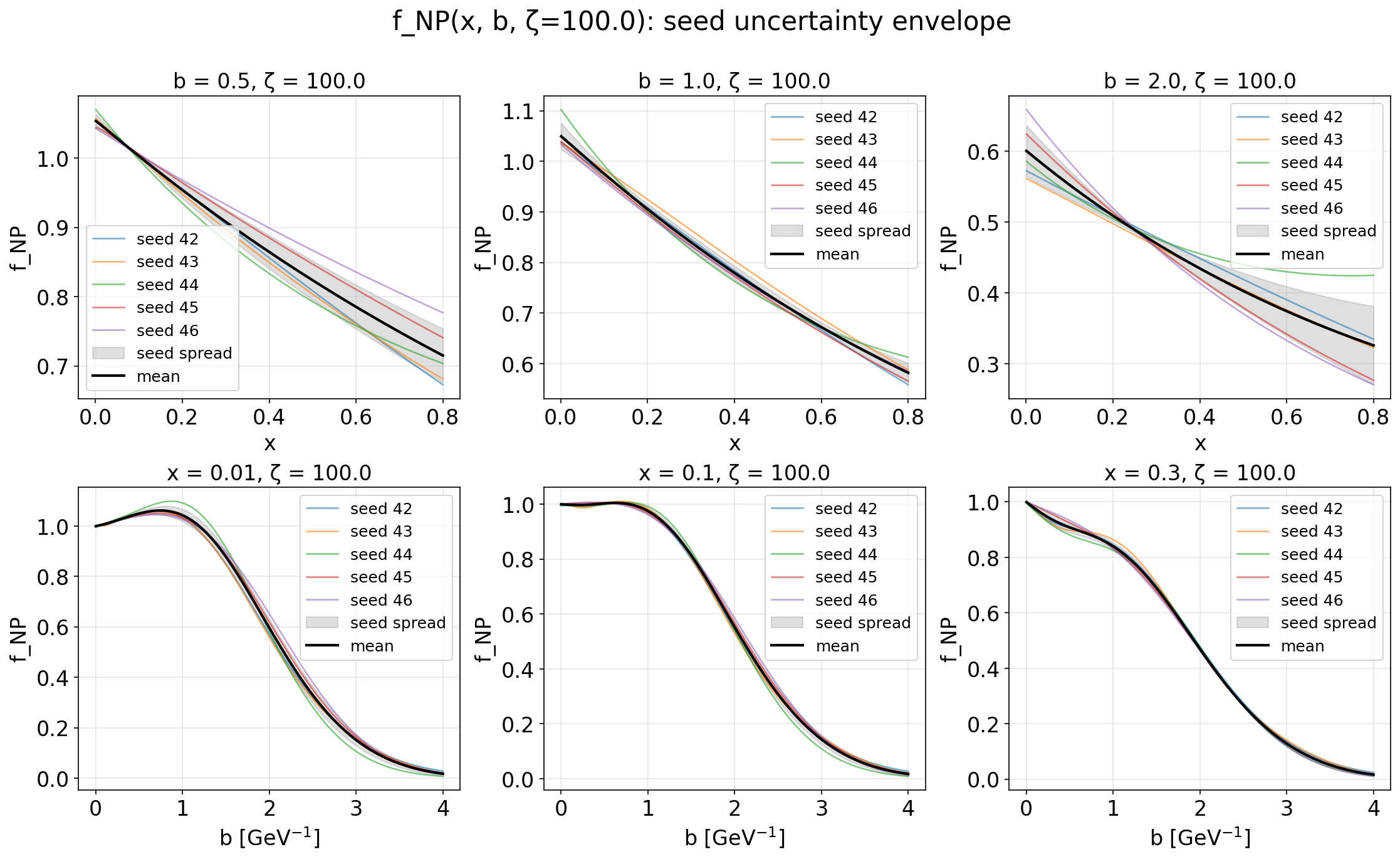}
    \caption{The non-perturbative $f_{\rm NP}(x,b_T;\zeta)$ as a function of $x$ (top row, representative $b_T$) and of $b_T$ (bottom row, representative $x$), evaluated at a representative rapidity scale $\zeta=100~\mathrm{GeV}^2$ with fixed reference $Q_0^2=1~\mathrm{GeV}^2$. 
    The function is evaluated from each neural-network seed (thin lines) together with the seed-ensemble mean and $\pm1\sigma$ band (shaded).
    }
    \label{fig:seed_ensemble_fnp}
\end{figure*}

The symbolic regression pipeline was executed across multiple independent NN seeds, each initialized with different random weights and trained independently on the experimental $\chi^2$. The resulting ensemble of extracted non-perturbative functions is shown in Fig.~\ref{fig:seed_ensemble_fnp}, displayed as a function of $x$ (at representative $b_T$) and of $b_T$ (at representative $x$) at a representative rapidity scale $\zeta = 100~\mathrm{GeV}^2$, with fixed reference $Q_0^2 = 1~\mathrm{GeV}^2$.

The spread across seeds provides a direct measure of the robustness of the extraction: where the band is narrow the recovered functional form is stable across different training initializations; where it is wider the NN training finds genuinely distinct solutions of comparable quality. This seed-to-seed variation arises from the stochastic nature of NN optimization combined with NN underdetermination in the Sudakov-suppressed region at large $b_T$ (see Sec.~\ref{sec:discussion}): the $\chi^2$ gradient signal reaching the shape networks is proportional to the Collins-Soper evolution factor, which is exponentially small for $b_T \gtrsim 4\ \mathrm{GeV}^{-1}$, leaving the networks poorly constrained there and sensitive to initialization.

%
%
From this ensemble we report in the next subsection the single formula that achieves the lowest worst-data subset $\chi^2/N_{s} < 3$ on the Pareto front - while being globally statistically compatible with 1 - in the (complexity, worst-data-subset $\chi^2$ proxy) plane, occurring at total complexity~$31$.  

The Pareto point in the (total complexity, worst-bin $\chi^2/\text{ndf}$) plane which meets the prior criteria occurs at complexity~$31$, where the worst-bin $\chi^2/\text{ndf}$ drops sharply from values above~$3.8$ at complexity~$28$ to $3.15$ at complexity~$29$ to~$2.96$ at complexity~$31$, and then plateaus  as complexity continues to grow above 31 (the worst-bin $\chi^2/\text{ndf}$ improves by only $0.09$ over those $23$ additional complexity units). We adopt 31 as the smallest complexity at which every experiment satisfies $\chi^2/N_s < 3$ and the worst-experiment plateau is established.
The selected formula --- the complexity-$31$ Pareto candidate --- achieves global $\chi^2/\text{ndf} \simeq 1.040$ ($\text{ndf} = 473$, $1\sigma$ interval $[0.975, 1.105]$ containing unity) and worst-data-subset $\chi^2/N_{s} \simeq 2.96$, with~$9$ free numerical constants distributed across $h_x$ (2), $h_b$ (2), $h_{xb}$ (2), and $g_2$ (3). The worst-data-subset value matches the parent NN worst-bin $\chi^2/N_{s}$ of~$2.89$ to within~$2.5\%$, indicating that the distillation introduces negligible degradation on the hardest-to-fit data.

Further increasing the complexity offers only marginal benefit: the worst-data-subset $\chi^2$ decreases by just $0.09$ over the next $23$ complexity units.

\subsection{Selected Formula}
\label{sec:results:selected_formula}
Symbolic regression applied to the trained factorized NN yields the following analytical expressions for the non-perturbative components, where $s = 1\ \mathrm{GeV}^{-1}$ is a fixed reference scale introduced to render the arguments of the non-linear functions dimensionless; its numerical value leaves all fitted constants unchanged:
\begin{align}
    h_x(x)        &= -0.854 - 1.45\,x \label{e:best_hx}\\
    h_b(b_T)      &= \frac{b_T}{s}\left(0.919 - 0.368\,\sqrt{b_T/s}\right) \label{e:hb}\\
    h_{xb}(x,b_T) &= \frac{b_T}{s}\left(0.766 - 0.0567\,(b_T/s - x)^2\right) \label{e:hxb}\\
    g_2(b_T)      &= 0.169 + 0.0328\,\sqrt{b_T/s + 0.636}
    \label{e:best_g2}
\end{align}
The complete nonperturbative function is therefore
\begin{multline*}
f_{\rm NP}(x, b_T, \zeta) = \exp\Bigl[-g_2^2(b_T) \, b_T^2 \,\ln(\zeta/Q_0^2)\Bigr] \\
\exp\Bigl[h_x(x) \, h_b(b_T) + h_{xb}(x, b_T)\Bigr]
\end{multline*}
with $Q_0^2 = 1~\text{GeV}^2$. The formula has total expression complexity~$31$ (summed across the four components), contains~$9$ free numerical constants,

The boundary condition $R(x, b_T\!=\!0) = 1$ is preserved exactly: both $h_b$ and $h_{xb}$ are at least linear in $b_T$,
contain explicit $(\cdot\,b)$ multipliers, 
and the first exponent vanishes through its $b_T^2$ factor. The formula contains no piecewise (absolute-value) operators; all four components are smooth analytic functions.

The numerical constants in Eqs.~\eqref{e:best_hx}-\eqref{e:best_g2} are quoted to three significant figures. A per-constant precision scan in which each constant is rounded in isolation to $k$ significant figures shows that no individual constant shifts the global $\chi^2/\text{ndf}$ by more than $10^{-3}$ at $k = 3$.

The $x$-dependent factor $h_x(x)$ is a simple linear function, negative across the entire physical region. The $b_T$-dependent factor
$h_b(b_T)$ grows approximately linearly at small $b_T$, reaches a maximum of about $0.85$ at $b_T/s\simeq2.77$, and then decreases because of the negative $\sqrt{b_T/s}$ term.
The cross term $h_{xb}(x, b_T)$ contains a $x$--$b_T$ correlation through the combination $b_T/s-x$:
at fixed $b_T$, it is an inverted parabola in~$x$, peaked along the diagonal $x \simeq b_T$ and decreasing quadratically away from this line. 
The Collins--Soper coupling $g_2(b_T)$ grows monotonically from $g_2(0) \simeq 0.195$ to $g_2(2.5~\text{GeV}^{-1}) \simeq 0.227$. The value at $b_T = 0$ closely matches the scalar $g_2$ extracted in the MAPNN fit~\cite{Bacchetta:2025ara}, and the slow monotonic growth captures a mild non-perturbative running of the Collins--Soper kernel.

\subsection{Fit Quality}
A breakdown of the reduced $\chi^2$ by experimental subset is given in Table~\ref{tab:chi2}. The fit shows good aggregate agreement across all experimental subsets. The global $\chi^2/\text{ndf} \simeq 1.040 \pm 0.065$ is statistically compatible with a reduced $\chi^2$ of 1.
The high-statistics LHC measurements are generally well described; the CMS data set in particular gives the smallest average subset-normalized value,
$\chi_s^2/N_s = 0.429$, compared with LHCb at $1.074$ and ATLAS at $1.274$. 
The CMS spectra are visually well reproduced across rapidity bins, with residual pulls mostly within the expected statistical range.
Tevatron and RHIC data sit close to unity ($0.864$ and $1.101$ respectively), and fixed-target Drell--Yan data averages to $1.179$, slightly above the global mean.

\begin{table}[h]
\centering
\begin{tabular}{lrrr}
\hline
Experiment & $N_{\rm dat}$ & $\bar{\chi}^2$ NN & $\bar{\chi}^2$ Symbolic \\
\hline
Fermilab  &  233 & $1.057$ & $1.179$ \\
RHIC          &    7 & $1.070$ & $1.101$ \\
Tevatron      &   71 & $0.803$ & $0.864$ \\
LHCb          &   21 & $1.083$ & $1.074$ \\
CMS           &   78 & $0.420$ & $0.429$ \\
ATLAS         &   72 & $1.213$ & $1.274$ \\
\hline
Total         &  482 & $0.941$ & 1.040 \\ 
\hline
\end{tabular}
\caption{Reduced $\bar\chi^2$ for each experimental subset, comparing the trained neural network of Ref.~\cite{Bacchetta:2025ara} and the selected complexity-31 symbolic formula. Both are evaluated on the same 482 data points using the full experimental covariance matrix, with correlated systematic uncertainties treated via analytic nuisance-parameter profiling identical to that of Ref.~\cite{Bacchetta:2025ara}. 
}
\label{tab:chi2}
\end{table}

\begin{figure}[htbp]
    \centering
    \includegraphics[width=0.48\textwidth]{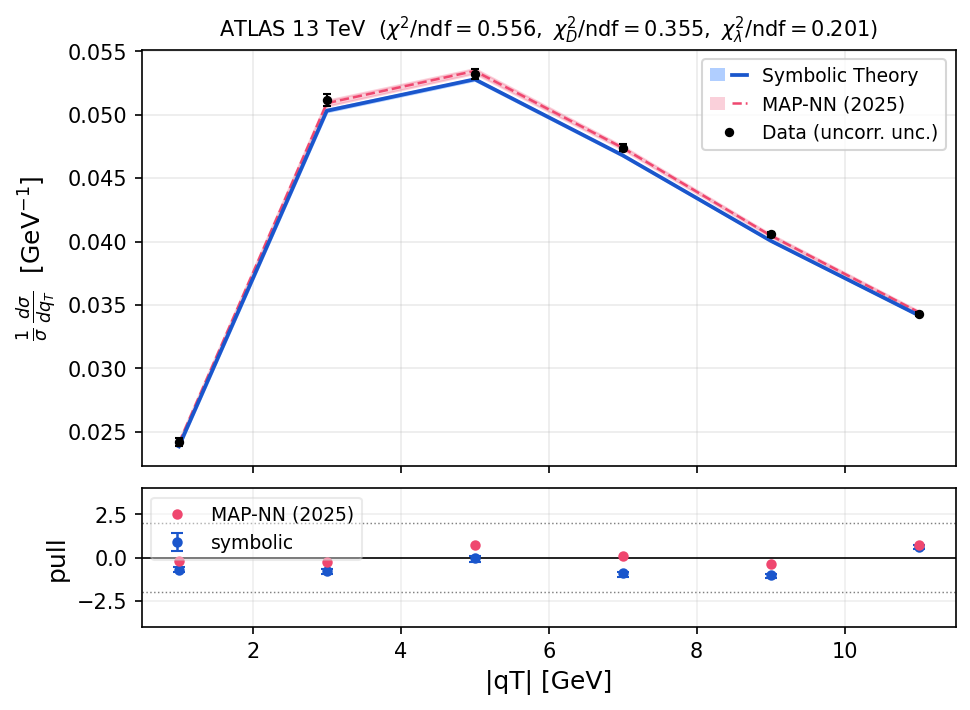}\\[0.5em]
    \includegraphics[width=0.48\textwidth]{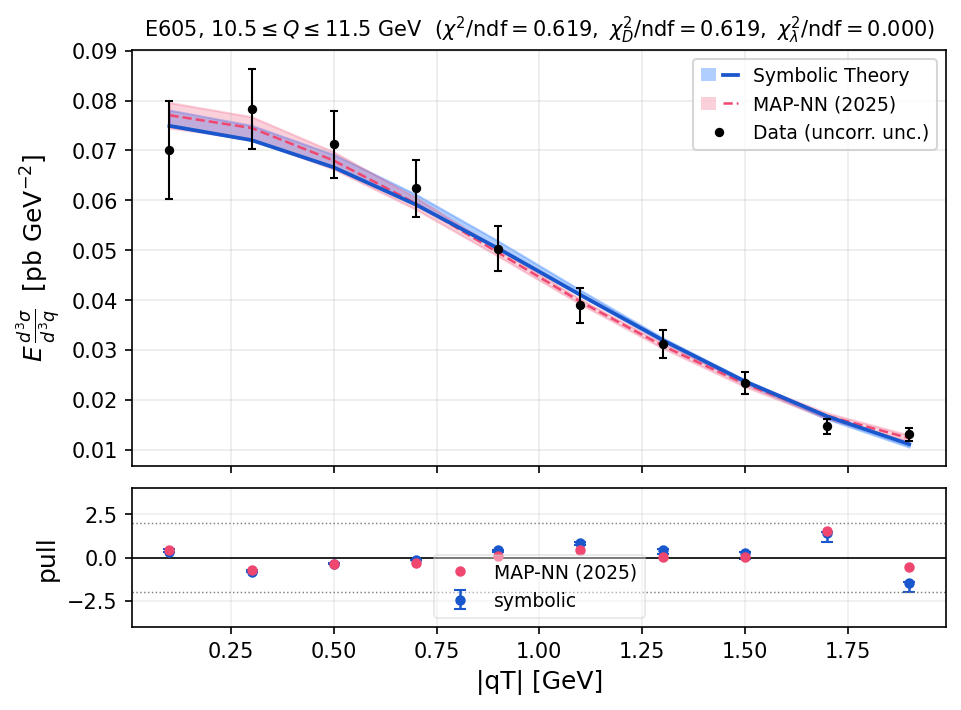}
    \caption{Drell-Yan cross section as a function of $q_T$.
    Top panel: ATLAS measurements at $\sqrt{s} = 13$~TeV and 
    $66 < Q < 116$~GeV bin~\cite{ATLAS3}. 
    Bottom panel: E605 data at $\sqrt{s} = 38.8$~GeV and $10.5 < Q < 11.5$~GeV bin~\cite{E605}. Each panel with its total ($\chi^2/\mathrm{ndf}$), uncorrelated ($\chi^2_D/\mathrm{ndf}$) and correlated-systematic ($\chi^2_\lambda/\mathrm{ndf}$) reduced $\chi^2$. Solid line for the result of this work from symbolic $f_{\rm NP}$, blue band spanning the Pareto-optimal symbolic candidates with global $\chi^2/\mathrm{ndf}$ statistically compatible with 1. Dashed line for the fit of Ref.~\cite{Bacchetta:2025ara} with 68\% uncertainty red band. Lower box in each panel for the post-fit pulls, (theory $-$ data)/$\sigma$, for the symbolic formula (blue) versus MAPNN (red).
    }
    \label{fig:representative_experiments}
\end{figure}

The fits and corresponding figures for all experimental data subsets are provided in the interactive Pareto explorer~\cite{pareto_explorer} for the selected solution with complexity $31$ and global $\chi^2/\mathrm{ndf}=1.040$.
Figure~\ref{fig:representative_experiments} shows representative comparisons of the $q_T$ distribution extracted from the symbolic $f_{\rm NP}$ against the ATLAS measurement at $\sqrt{s}=13~\text{TeV}$ and $66 < Q < 116$ GeV bin, and against E605 at $\sqrt{s}=38.8~\text{GeV}$ and $10.5 < Q < 11.5~\text{GeV}$ bin. The symbolic $f_{\rm NP}$ matches also the MAPNN  benchmark fit~\cite{Bacchetta:2025ara} for the two illustrative datasets shown. Panel titles report the total, uncorrelated, and correlated-systematic reduced $\chi^2$ ($\chi^2/\mathrm{ndf}$, $\chi^2_D/\mathrm{ndf}$, $\chi^2_\lambda/\mathrm{ndf}$) respectively; the red band is the MAPNN benchmark of Ref.~\cite{Bacchetta:2025ara} with its replica uncertainty, the blue band spans the Pareto-optimal symbolic candidates with global $\chi^2/\mathrm{ndf}$ statistically compatible with 1, and the lower box shows the post-fit pulls, (theory $-$ data)/$\sigma$, for the symbolic formula (blue) versus MAPNN (red).

 The fit shows good aggregate agreement across all experimental subsets.
Within these aggregate numbers, a small set of individual measurements show notable tension. The largest contributions are from E772 at $8<Q<9~\text{GeV}$ (the worst-fit experiment, with a subset reduced $\chi^2 \simeq 2.96$), ATLAS at $7~\text{TeV}$ and $|y| < 1$ ($\chi^2 \simeq 2.70$), E605 at $8<Q<9~\text{GeV}$ ($\chi^2 \simeq 2.65$), E772 at $7<Q<8~\text{GeV}$ ($\chi^2 \simeq 2.55$), and D0 Run~II ($\chi^2 \simeq 2.49$). 
The tensions concentrate in two phase-space regions: (i)~low-$Q$ fixed-target Drell--Yan (E772 and E605), and (ii)~central-rapidity hadron-collider bins (ATLAS at $7~\text{TeV}$ and $|y|<1$, and D0 Run~II). All remaining data subsets fall below a subset $\chi^2 \simeq 2.5$, with the large majority below~$2$.

The selected $h_b(b_T)$ develops a node in the $b_T$-dependent exponent at $b_T \simeq 6.24~\mathrm{GeV}^{-1}$. This zero crossing lies outside the range constrained by the symbolic-regression fit ($b_T < b_T^{\rm cut} = 2.5~\mathrm{GeV}^{-1}$; see Sec.~\ref{ss:symbolic_regression}) and is therefore an extrapolation beyond the data-supported region rather than a data-driven feature.

%% file: 6_discussion.tex
\section{Discussion}\label{sec:discussion}

\subsection{Physical Interpretation of the Selected Formula}

The components of the selected formula, presented in Sec.~\ref{sec:results:selected_formula}, admit partial physical interpretation; below we discuss what can be connected to known QCD physics and what remains a compact but purely data-driven representation.

\paragraph{The $x$-dependent factor.}
The $x$-dependent factor $h_x(x) = -0.854 - 1.45 \, x$ is a linear, monotonically decreasing function that remains strictly negative on the physical support $x \in [0,\,1]$, ranging from $-0.854$ at $x = 0$ to $-2.306$ at $x = 1$. Unlike earlier candidate formulas with a sign change in $x$, the selected expression places the zero of $h_x$ outside the physical range, so the factorized component $h_x \, h_b$ has a uniform sign across all $x$. The $x$-dependence of $R(x,b_T)$ then enters only as a smooth modulation of the amplitude of the $b_T$-space suppression, consistent with the data-driven preference for moderate, slowly-varying $x$-dependence.

\paragraph{The $b_T$-dependent factor.}
The $b_T$-dependent factor $h_b(b_T) = b_T/s\, \left(0.919 - 0.368\,\sqrt{b_T/s}\right)$ vanishes exactly at $b_T = 0$ as required by the boundary condition $R(x,0) = 1$; this is guaranteed analytically by the overall factor of $b_T$ rather than imposed only approximately by training. The function is positive on $0 < b_T < 6.24\ \mathrm{GeV}^{-1}$, encompassing the extraction domain $b_T < b_T^\mathrm{cut} = 2.5\ \mathrm{GeV}^{-1}$, and peaks at $b_T \approx 2.77\ \mathrm{GeV}^{-1}$ with maximum value $\approx 0.85$. The data-supported range is estimated as $b_T \lesssim 4.5\ \mathrm{GeV}^{-1}$: for $g_2 \approx 0.19$ and the minimum rapidity scale $\zeta \sim 25\ \mathrm{GeV}^2$ in our dataset, the data contribution falls below $5\%$ at $b_T \simeq 4.5\ \mathrm{GeV}^{-1}$, beyond which data carry negligible constraint. Combined with the strictly negative $h_x$, the product $h_x h_b$ is negative within the extraction domain $b_T < 2.5\ \mathrm{GeV}^{-1}$, where $h_b$ remains positive. Beyond the extraction domain, $h_b$ develops a node at $b_T \simeq 6.24\ \mathrm{GeV}^{-1}$ where the factorized product changes sign; this is an extrapolation artifact discussed in \ref{sec:limitations}. The leading small-$b_T$ behavior is $h_b(b_T) \simeq 0.919\,b_T$, \textit{i.e.}, linear rather than quadratic. 

\paragraph{The cross term.}
The cross term $h_{xb}(x, b_T) = b_T/s\, \left(0.766 - 0.0567\,(b_T/s - x)^2\right)$ contains a leading $0.766\,b_T$ contribution comparable in magnitude to the factorized product $h_x h_b$ throughout the data-supported region. At  $(x,b_T)=(0.1, 1.0\ \mathrm{GeV}^{-1})$ the cross term contributes $+0.720$ to the exponent while $h_x(0.1),h_b(1.0) = -0.550$, so $h_{xb}$ partially cancels the factorized suppression and substantially reshapes $R(x,b_T)$. At small $x \lesssim 0.1$, the cross-term contribution is large enough that $f_\mathrm{NP}$ briefly exceeds 1 at moderate $b_T \sim 0.5$--$1\ \mathrm{GeV}^{-1}$; this is not excluded by the CSS formalism, which requires only $f_\mathrm{NP} \to 1$ at $b_T = 0$ and $f_\mathrm{NP} \to 0$ at $b_T \to \infty$, the latter being guaranteed by the Sudakov factor independently of $R$. The $(b_T/s-x)^2$ modulation provides the only 
$x$--$b_T$ correlation in the formula, controlling the alignment between the small-$x$ / small-$b_T$ region (where the modulation is suppressed) and the small-$x$ / large-$b_T$ region (where the quadratic term grows). The boundary condition $h_{xb}(x,0)=0$ is again preserved exactly by the explicit factor of $b_T$.

\paragraph{The \it Collins-Soper kernel.}
The Collins-Soper kernel contains the $g_2(b_T)$ function. The best result is given by $g_2(b_T) = 0.169 + 0.0328\,\sqrt{b_T/s + 0.636}$; it is a mildly $b_T$-dependent function that varies slowly from $g_2 \simeq 0.195~\text{GeV}$ at $b_T = 0$ to $g_2 \simeq 0.247~\text{GeV}$ at $b_T = 5~\text{GeV}^{-1}$. The small-$b_T$ value is consistent with prior global fits~\cite{Bacchetta:2022awv,Bacchetta:2025ara}. The selected functional form prefers a weak but nonzero $b_T$-dependence over a constant; the residual variation across the data-supported range remains modest, indicating that current Drell--Yan data constrain $g_2$ to a narrow window rather than resolving any rich $b_T$-dependence.

\subsection{Limitations and Future Work}
\label{sec:limitations}

Several limitations affect the interpretation of this result and motivate future work. First, and most importantly, the result is conditioned on the specific functional parameterization adopted
in
Eqs.~(8)--(9): writing $R(x,b_T)=\exp[h_x(x)\,h_b(b_T)+ h_{xb}(x,b_T)]$ is an architectural choice, and the interpretability of the extraction rests on it. This choice
is what reduces an intractable 2D symbolic-regression problem to two 1D problems plus a small cross term; fitting
$R(x,b_T)$ as a single unconstrained 2D network yields no candidate with $\chi^2/N_{\rm dat}<1.3$ and no compactly distillable expression. Keeping the cross term $h_{xb}$ small enough to distill, in turn, relies on an explicit sparsity prior ($L_1$ penalty) together with a hard capacity cap (a single four-neuron hidden layer) and subset-balanced training; these ingredients bias the solution toward separability, so the smallness of $h_{xb}$ is in part imposed rather than purely data-driven. The extracted formula should therefore be read as the most compact description \emph{within this ansatz}, and alternative enveloping structures may yield qualitatively different component functions of comparable fit quality. Relaxing these choices --- higher-capacity or unregularized cross terms, or entirely different factorization templates --- while retaining distillability is an important direction for future work.

Second, the analysis assumes a flavor-universal $f_{\rm NP}$, an approximation shared with the reference fit~\cite{Bacchetta:2025ara}; flavor dependence enters the full TMD only through perturbative matching onto collinear PDFs, and a flavor-dependent $f_{\rm NP}$ would require additional constraints, e.g.\ from semi-inclusive DIS, to disentangle~\cite{Bacchetta:2024qr}. Third, the neural network is trained against a single reference determination rather than an ensemble of fits; the seed-to-seed spread of Fig.~2 captures optimization variance but not the full theoretical and experimental uncertainty of the underlying extraction. A natural extension is to train directly on replica ensembles, propagating experimental and model uncertainty into the symbolic constants. Fourth, the extracted formulas are constrained only within $b_T<b_T^{\rm cut}=2.5~\mathrm{GeV}^{-1}$; features outside this range, such as the $h_b$ node at $b_T\simeq6.24~\mathrm{GeV}^{-1}$, are extrapolations. Future directions include replica-based uncertainty quantification, tests of flavor dependence with SIDIS data, extension to polarized TMDs, and validation of the discovered structures in independent processes.

%% file: 7_conclusion.tex
\section{Conclusion}

We have presented the extraction of an analytical formula for the nonperturbative component of the TMD from a data-driven neural-network (NN) fit, using symbolic regression to discover compact expressions directly from the learned network behavior. Starting from a factorized NN parametrization at N$^3$LL accuracy, trained against Drell--Yan data from fixed-target experiments, the Tevatron, RHIC, and the LHC, we applied PySR to each network component independently and selected the final formula by evaluating all combinations on the full experimental $\chi^2$ via a Pareto analysis in the
(complexity, $\chi^2/\text{ndf}$) plane.

The obtained formula achieves $\chi^2/\text{ndf} = 1.040 \pm 0.065$ over $482$ data points, comparable to the parent NN fit with $\chi^2/\text{ndf} = 0.941$, while containing only $9$ free numerical constants and total expression complexity~$31$. The selected expression includes a non-trivial cross term, $h_{xb}(x,b_T)$,
that contributes at leading order alongside the factorized backbone $h_x \, h_b$, indicating that the data prefer a structure with mild but genuine $x$--$b_T$ correlation rather than strict separability.  

This work demonstrates that symbolic regression can extract interpretable analytical expressions from NN fits, offering a complement to purely data-driven machine-learning approaches. The combination of a factorized NN architecture, direct experimental $\chi^2$ training, and Pareto-guided symbolic extraction provides one systematic pathway from data to formula in which the individual component functions $h_x$, $h_b$, $h_{xb}$, and $g_2$ are discovered rather than postulated. The overall structure --- a 
 Collins--Soper evolution factor multiplying an exponential of a factorized backbone plus additive cross-term --- is itself an architectural choice, and the boundary conditions $R(x,0) = 1$ and $h_{xb}(x,0) = 0$ are enforced by construction. Alternative structures may yield qualitatively different component functions with comparable fit quality. Extensions of this work include rigorous uncertainty quantification via Markov-chain Monte Carlo, tests of flavor dependence using semi-inclusive DIS data, application to polarized TMDs, and validation of the extracted features in related processes.

%% file: acknowledgments.tex
\vspace{0.8cm}
\section*{Acknowledgments}

This material is based upon work supported by the National Science Foundation under Grant No.~2443510, an NSF CAREER award.
The William \& Mary group acknowledges support from this grant.
The authors acknowledge W\&M  Research Computing for providing computational resources and technical support that have contributed to the results reported within this article.
%
%
CB contribution is based upon work supported by Laboratory Directed Research and
Development (LDRD) funding from Argonne National Laboratory, provided by the Director, Office of Science, of
the U.S. Department of Energy under Contract No. DE-AC02-06CH11357. The work of V.B. and M.C. has
been supported by l’Agence Nationale de la Recherche (ANR), project ANR-24-CE31-7061-01.

%% file: apssamp.bib
@PREAMBLE{
 "\providecommand{\noopsort}[1]{}" 
 # "\providecommand{\singleletter}[1]{#1}%" 
}

@article{Bacchetta:2024yzl,
    author = "Bacchetta, Alessandro and Bongallino, Alessia and Cerutti, Matteo and Radici, Marco and Rossi, Lorenzo",
    collaboration = "MAP (Multi-dimensional Analyses of Partonic distributions)",
    title = "{Exploring the Three-Dimensional Momentum Distribution of Longitudinally Polarized Quarks in the Proton}",
    eprint = "2409.18078",
    archivePrefix = "arXiv",
    primaryClass = "hep-ph",
    reportNumber = "JLAB-THY-24-4204",
    doi = "10.1103/PhysRevLett.134.121901",
    journal = "Phys. Rev. Lett.",
    volume = "134",
    number = "12",
    pages = "121901",
    year = "2025"
}

@article{Rossi:2025pwh,
    author = "Rossi, Lorenzo and Bacchetta, Alessandro and Cerutti, Matteo and Radici, Marco",
    collaboration = "MAP (Multi-dimensional Analyses of Partonic distributions)",
    title = "{New insights from the flavor dependence of quark transverse momentum distributions in the pion}",
    eprint = "2509.25098",
    archivePrefix = "arXiv",
    primaryClass = "hep-ph",
    doi = "10.1016/j.physletb.2026.140482",
    journal = "Phys. Lett. B",
    volume = "877",
    pages = "140482",
    year = "2026"
}

@article{Cerutti:2022lmb,
    author = "Cerutti, Matteo and Rossi, Lorenzo and Venturini, Simone and Bacchetta, Alessandro and Bertone, Valerio and Bissolotti, Chiara and Radici, Marco",
    collaboration = "MAP (Multi-dimensional Analyses of Partonic distributions)",
    title = "{Extraction of pion transverse momentum distributions from Drell-Yan data}",
    eprint = "2210.01733",
    archivePrefix = "arXiv",
    primaryClass = "hep-ph",
    doi = "10.1103/PhysRevD.107.014014",
    journal = "Phys. Rev. D",
    volume = "107",
    number = "1",
    pages = "014014",
    year = "2023"
}

@article{Bacchetta:2025ara,
    author = "Bacchetta, Alessandro and Bertone, Valerio and Bissolotti, Chiara and Cerutti, Matteo and Radici, Marco and Rodini, Simone and Rossi, Lorenzo",
    collaboration = "MAP (Multi-dimensional Analyses of Partonic distributions)",
    title = "{Neural-Network Extraction of Unpolarized Transverse-Momentum-Dependent Distributions}",
    eprint = "2502.04166",
    archivePrefix = "arXiv",
    primaryClass = "hep-ph",
    reportNumber = "DESY-25-022, JLAB-THY-25-4221",
    doi = "10.1103/csc2-bj91",
    journal = "Phys. Rev. Lett.",
    volume = "135",
    number = "2",
    pages = "021904",
    year = "2025"
}

@article{Moos:2023yfa,
    author = "Moos, Valentin and Scimemi, Ignazio and Vladimirov, Alexey and Zurita, Pia",
    title = "{Extraction of unpolarized transverse momentum distributions from the fit of Drell-Yan data at N$^{4}$LL}",
    eprint = "2305.07473",
    archivePrefix = "arXiv",
    primaryClass = "hep-ph",
    reportNumber = "IPARCOS-UCM-035",
    doi = "10.1007/JHEP05(2024)036",
    journal = "JHEP",
    volume = "05",
    pages = "036",
    year = "2024"
}

@article{Bacchetta:2022awv,
    author = "Bacchetta, Alessandro and Bertone, Valerio and Bissolotti, Chiara and Bozzi, Giuseppe and Cerutti, Matteo and Piacenza, Fulvio and Radici, Marco and Signori, Andrea",
    collaboration = "MAP (Multi-dimensional Analyses of Partonic distributions)",
    title = "{Unpolarized transverse momentum distributions from a global fit of Drell-Yan and semi-inclusive deep-inelastic scattering data}",
    eprint = "2206.07598",
    archivePrefix = "arXiv",
    primaryClass = "hep-ph",
    doi = "10.1007/JHEP10(2022)127",
    journal = "JHEP",
    volume = "10",
    pages = "127",
    year = "2022"
}

@article{Bury:2022czx,
    author = "Bury, Marcin and Hautmann, Francesco and Leal-Gomez, Sergio and Scimemi, Ignazio and Vladimirov, Alexey and Zurita, Pia",
    title = "{PDF bias and flavor dependence in TMD distributions}",
    eprint = "2201.07114",
    archivePrefix = "arXiv",
    primaryClass = "hep-ph",
    reportNumber = "UWThPh 2021-29, CERN-TH-2022-126",
    doi = "10.1007/JHEP10(2022)118",
    journal = "JHEP",
    volume = "10",
    pages = "118",
    year = "2022"
}

@article{Bacchetta:2019sam,
    author = "Bacchetta, Alessandro and Bertone, Valerio and Bissolotti, Chiara and Bozzi, Giuseppe and Delcarro, Filippo and Piacenza, Fulvio and Radici, Marco",
    title = "{Transverse-momentum-dependent parton distributions up to N$^{3}$LL from Drell-Yan data}",
    eprint = "1912.07550",
    archivePrefix = "arXiv",
    primaryClass = "hep-ph",
    reportNumber = "JLAB-THY-19-3121",
    doi = "10.1007/JHEP07(2020)117",
    journal = "JHEP",
    volume = "07",
    pages = "117",
    year = "2020"
}

@article{Scimemi:2019cmh,
    author = "Scimemi, Ignazio and Vladimirov, Alexey",
    title = "{Non-perturbative structure of semi-inclusive deep-inelastic and Drell-Yan scattering at small transverse momentum}",
    eprint = "1912.06532",
    archivePrefix = "arXiv",
    primaryClass = "hep-ph",
    doi = "10.1007/JHEP06(2020)137",
    journal = "JHEP",
    volume = "06",
    pages = "137",
    year = "2020"
}

@article{Bertone:2019nxa,
    author = "Bertone, Valerio and Scimemi, Ignazio and Vladimirov, Alexey",
    title = "{Extraction of unpolarized quark transverse momentum dependent parton distributions from Drell-Yan/Z-boson production}",
    eprint = "1902.08474",
    archivePrefix = "arXiv",
    primaryClass = "hep-ph",
    doi = "10.1007/JHEP06(2019)028",
    journal = "JHEP",
    volume = "06",
    pages = "028",
    year = "2019"
}

@article{Scimemi:2017etj,
    author = "Scimemi, Ignazio and Vladimirov, Alexey",
    title = "{Analysis of vector boson production within TMD factorization}",
    eprint = "1706.01473",
    archivePrefix = "arXiv",
    primaryClass = "hep-ph",
    doi = "10.1140/epjc/s10052-018-5557-y",
    journal = "Eur. Phys. J. C",
    volume = "78",
    number = "2",
    pages = "89",
    year = "2018"
}

@article{Bacchetta:2017gcc,
    author = "Bacchetta, Alessandro and Delcarro, Filippo and Pisano, Cristian and Radici, Marco and Signori, Andrea",
    title = "{Extraction of partonic transverse momentum distributions from semi-inclusive deep-inelastic scattering, Drell-Yan and Z-boson production}",
    eprint = "1703.10157",
    archivePrefix = "arXiv",
    primaryClass = "hep-ph",
    reportNumber = "JLAB-THY-17-2437",
    doi = "10.1007/JHEP06(2017)081",
    journal = "JHEP",
    volume = "06",
    pages = "081",
    year = "2017",
    note = "[Erratum: JHEP 06, 051 (2019)]"
}

@article{Boussarie:2023izj,
    author = "Boussarie, Renaud and others",
    title = "{TMD Handbook}",
    eprint = "2304.03302",
    archivePrefix = "arXiv",
    primaryClass = "hep-ph",
    reportNumber = "JLAB-THY-23-3780, LA-UR-21-20798, MIT-CTP/5386",
    year = "2023",
    month = "4",
    journal = ""
}

@article{cranmer2023interpretable,
  title={{Interpretable machine learning for science with \texttt{PySR} and \texttt{SymbolicRegression.jl}}},
  author={Cranmer, Miles},
  journal={arXiv preprint arXiv:2305.01582},
  year={2023}
}

@misc{cranmer2020discoveringsymbolicmodelsdeep,
      title={{Discovering Symbolic Models from Deep Learning with Inductive Biases}}, 
      author={Miles Cranmer and Alvaro Sanchez-Gonzalez and Peter Battaglia and Rui Xu and Kyle Cranmer and David Spergel and Shirley Ho},
      year={2020},
      eprint={2006.11287},
      archivePrefix={arXiv},
      primaryClass={cs.LG},
      url={https://arxiv.org/abs/2006.11287}, 
}

@book{Collins2011,
      author = "Collins, John",
      title = "{Foundations of perturbative QCD}",
      publisher = "Cambridge Univ. Press",
      address = "New York, NY",
      series = "Cambridge monographs on particle physics, nuclear physics and cosmology",
      year = "2011",
      url = "https://cds.cern.ch/record/1350496",
      doi = "10.1017/CBO9780511975592",
}

@article{Collins1985,
title = {{Transverse momentum distribution in Drell-Yan pair and W and Z boson production}},
journal = {Nuclear Physics B},
volume = {250},
number = {1},
pages = {199-224},
year = {1985},
issn = {0550-3213},
doi = {https://doi.org/10.1016/0550-3213(85)90479-1},
url = {https://www.sciencedirect.com/science/article/pii/0550321385904791},
author = {J.C. Collins and Davison E. Soper and George Sterman}
}

@software{NangaParbat,
  title={{Nanga Parbat: A Fitting Framework for the Determination of the Non-perturbative Component of TMD Distributions}},
  author={Bertone, Valerio and Bacchetta, Alessandro and Bissolotti, Chiara and Cerutti, Matteo and Radici, Marco and Rodini, Simone and Rossi, Lorenzo},
  year={2024},
  url={https://github.com/MapCollaboration/NangaParbat},
  version={1.5.0}
}

@article{Anedda:2026cox,
    author = "Anedda, Simone and Bertone, Valerio and Bozzi, Giuseppe and Cerutti, Matteo",
    title = "{A first extraction of gluon TMDs from Higgs data at the LHC}",
    eprint = "2605.28216",
    archivePrefix = "arXiv",
    primaryClass = "hep-ph",
    month = "5",
    year = "2026",
    journal = ""
}

@article{Cerutti:2026apy,
    author = "Cerutti, Matteo and Simonelli, Andrea",
    title = "{The impact of prescriptions in phenomenological extractions of Transverse Momentum Dependent distributions}",
    eprint = "2603.19088",
    archivePrefix = "arXiv",
    primaryClass = "hep-ph",
    month = "3",
    year = "2026",
    journal = ""
}

@article{Bacchetta:2024qr,
   title={{Flavor dependence of unpolarized quark transverse momentum distributions from a global fit}},
   volume={2024},
   ISSN={1029-8479},
   url={http://dx.doi.org/10.1007/JHEP08(2024)232},
   DOI={10.1007/jhep08(2024)232},
   number={8},
   journal={Journal of High Energy Physics},
   publisher={Springer Science and Business Media LLC},
   author={Bacchetta, Alessandro and Bertone, Valerio and Bissolotti, Chiara and Bozzi, Giuseppe and Cerutti, Matteo and Delcarro, Filippo and Radici, Marco and Rossi, Lorenzo and Signori, Andrea},
   year={2024},
   month=Aug 
}

@article{ogata2005numerical,
  title={{A numerical integration formula based on the Bessel functions}},
  author={Ogata, Hidenori},
  journal={Publications of the Research Institute for Mathematical Sciences},
  volume={41},
  number={4},
  pages={949--970},
  year={2005},
  url={https://ems.press/content/serial-article-files/40928?nt=1}
}

@misc{pytorch,
      title={PyTorch: An Imperative Style, High-Performance Deep Learning Library}, 
      author={Adam Paszke and Sam Gross and Francisco Massa and Adam Lerer and James Bradbury and Gregory Chanan and Trevor Killeen and Zeming Lin and Natalia Gimelshein and Luca Antiga and Alban Desmaison and Andreas Köpf and Edward Yang and Zach DeVito and Martin Raison and Alykhan Tejani and Sasank Chilamkurthy and Benoit Steiner and Lu Fang and Junjie Bai and Soumith Chintala},
      year={2019},
      eprint={1912.01703},
      archivePrefix={arXiv},
      primaryClass={cs.LG},
      url={https://arxiv.org/abs/1912.01703}, 
}

@article{Angeles_Martinez_2015,
   title={Transverse Momentum Dependent (TMD) Parton Distribution Functions: Status and Prospects},
   volume={46},
   ISSN={1509-5770},
   url={http://dx.doi.org/10.5506/APhysPolB.46.2501},
   DOI={10.5506/aphyspolb.46.2501},
   number={12},
   journal={Acta Physica Polonica B},
   publisher={Jagiellonian University},
   author={Angeles-Martinez, R. and Bacchetta, A. and Balitsky, I.I. and Boer, D. and Boglione, M. and Boussarie, R. and Ceccopieri, F.A. and Cherednikov, I.O. and Connor, P. and Echevarria, M.G. and Ferrera, G. and Grados Luyando, J. and Hautmann, F. and Jung, H. and Kasemets, T. and Kutak, K. and Lansberg, J.P. and Lelek, A. and Lykasov, G. and Madrigal Martinez, J.D. and Mulders, P.J. and Nocera, E.R. and Petreska, E. and Pisano, C. and Plačakytė, R. and Radescu, V. and Radici, M. and Schnell, G. and Scimemi, I. and Signori, A. and Szymanowski, L. and Taheri Monfared, S. and Van der Veken, F.F. and van Haevermaet, H.J. and Van Mechelen, P. and Vladimirov, A.A. and Wallon, S.},
   year={2015},
   pages={2501} }

@article{E605,
  title = {Dimuon production in proton-copper collisions at $\sqrt{s}=38.8$ GeV},
  author = {Moreno, G. and Brown, C. N. and Cooper, W. E. and Finley, D. and Hsiung, Y. B. and Jonckheere, A. M. and Jostlein, H. and Kaplan, D. M. and Lederman, L. M. and Hemmi, Y. and Imai, K. and Miyake, K. and Nakamura, T. and Sasao, N. and Tamura, N. and Yoshida, T. and Maki, A. and Sakai, Y. and Gray, R. and Luk, K. B. and Rutherfoord, J. P. and Straub, P. B. and Williams, R. W. and Young, K. K. and Adams, M. R. and Glass, H. and Jaffe, D. and McCarthy, R. L. and Crittenden, J. A. and Smith, S. R.},
  journal = {Phys. Rev. D},
  volume = {43},
  issue = {9},
  pages = {2815--2835},
  numpages = {0},
  year = {1991},
  month = {May},
  publisher = {American Physical Society},
  doi = {10.1103/PhysRevD.43.2815},
  url = {https://link.aps.org/doi/10.1103/PhysRevD.43.2815}
}

@article{E228,
  author = {Ito, A. S. and Fisk, R. J. and J\"ostlein, H. and Kaplan, D. M. and Herb, S. W. and Hom, D. C. and Lederman, L. M. and Snyder, H. D. and Yoh, J. K. and Brown, B. C. and Brown, C. N. and Innes, W. R. and Kephart, R. D. and Ueno, K. and Yamanouchi, T.},
    title = {Measurement of the continuum of dimuons produced in high-energy proton-nucleus collisions},
  journal = {Phys. Rev. D},
  volume = {23},
  issue = {3},
  pages = {604--633},
  numpages = {0},
  year = {1981},
  month = {Feb},
  publisher = {American Physical Society},
  doi = {10.1103/PhysRevD.23.604},
  url = {https://link.aps.org/doi/10.1103/PhysRevD.23.604}
}

@article{E772,
  title = {Cross sections for the production of high-mass muon pairs from 800 GeV proton bombardment of $^{2}\mathrm{H}$},
  author = {McGaughey, P. L. and Moss, J. M. and Alde, D. M. and Baer, H. W. and Carey, T. A. and Garvey, G. T. and Klein, A. and Lee, C. and Leitch, M. J. and Lillberg, J. and Mishra, C. S. and Peng, J. C. and Brown, C. N. and Cooper, W. E. and Hsiung, Y. B. and Adams, M. R. and Guo, R. and Kaplan, D. M. and McCarthy, R. L. and Danner, G. and Wang, M. and Barlett, M. and Hoffmann, G.},
  journal = {Phys. Rev. D},
  volume = {50},
  issue = {5},
  pages = {3038--3045},
  numpages = {0},
  year = {1994},
  month = {Sep},
  publisher = {American Physical Society},
  doi = {10.1103/PhysRevD.50.3038},
  url = {https://link.aps.org/doi/10.1103/PhysRevD.50.3038}
}

@article{CDF1,
  title = {Transverse Momentum and Total Cross Section of ${\mathit{e}}^{+}{\mathit{e}}^{\ensuremath{-}}$ Pairs in the Z-Boson Region from $\mathit{p}\overline{\mathit{p}}$ Collisions at $\sqrt{s}=1.8\mathrm{TeV}$},
  author = {Affolder, T. and others},
  collaboration = {CDF Collaboration},
  journal = {Phys. Rev. Lett.},
  volume = {84},
  issue = {5},
  pages = {845--850},
  year = {2000},
  month = {Jan},
  publisher = {American Physical Society},
  doi = {10.1103/PhysRevLett.84.845},
  url = {https://link.aps.org/doi/10.1103/PhysRevLett.84.845}
}

@article{CDF2,
  title = {Transverse momentum cross section of ${e}^{+}{e}^{-}$ pairs in the $Z$-boson region from $p\overline{p}$ collisions at $\sqrt{s}=1.96\text{ }\mathrm{TeV}$},
  author = {Aaltonen, T. and others},
  collaboration = {CDF Collaboration},
  journal = {Phys. Rev. D},
  volume = {86},
  issue = {5},
  pages = {052010},
  year = {2012},
  month = {Sep},
  publisher = {American Physical Society},
  doi = {10.1103/PhysRevD.86.052010},
  url = {https://link.aps.org/doi/10.1103/PhysRevD.86.052010}
}

@article{D01,
  title = {Measurement of the inclusive differential cross section for Z bosons as a function of transverse momentum in $\overline{p}p$ collisions at $\sqrt{s}=1.8$ TeV},
  author = {Abbott, B. and others},
  collaboration = {D0 Collaboration},
  journal = {Phys. Rev. D},
  volume = {61},
  issue = {3},
  pages = {032004},
  year = {2000},
  month = {Jan},
  publisher = {American Physical Society},
  doi = {10.1103/PhysRevD.61.032004},
  url = {https://link.aps.org/doi/10.1103/PhysRevD.61.032004}
}

@article{D02,
  title = {Measurement of the Shape of the Boson-Transverse Momentum Distribution in $p\overline{p}\rightarrow Z/{\gamma}^{*}\rightarrow{e}^{+}{e}^{-}+X$ Events Produced at $\sqrt{s}=1.96\text{ }\mathrm{TeV}$},
  author = {Abazov, V. M. and others},
  collaboration = {D0 Collaboration},
  journal = {Phys. Rev. Lett.},
  volume = {100},
  issue = {10},
  pages = {102002},
  year = {2008},
  month = {Mar},
  publisher = {American Physical Society},
  doi = {10.1103/PhysRevLett.100.102002},
  url = {https://link.aps.org/doi/10.1103/PhysRevLett.100.102002}
}

@article{D03,
  author = {Abazov, Victor Mukhamedovich and others},
  collaboration = {D0 Collaboration},
  title = "{Measurement of the Normalized $Z/\gamma^* \to \mu^+\mu^-$ Transverse Momentum Distribution in $p\bar{p}$ Collisions at $\sqrt{s}=1.96$ TeV}",
  eprint = {1006.0618},
  archivePrefix = {arXiv},
  primaryClass = {hep-ex},
  reportNumber = {FERMILAB-PUB-10-183-E},
  doi = {10.1016/j.physletb.2010.09.012},
  journal = {Phys. Lett. B},
  volume = {693},
  pages = {522--530},
  year = {2010}
}

@article{STAR,
  title = {Measurements of the $Z^0/\gamma^*$ cross section and transverse single spin asymmetry in 510 GeV $p+p$ collisions},
  author = {Abdulhamid, M. I. and others},
  collaboration = {STAR Collaboration},
  journal = {Phys. Lett. B},
  volume = {854},
  pages = {138715},
  year = {2024},
  doi = {10.1016/j.physletb.2024.138715},
  url = {https://www.sciencedirect.com/science/article/pii/S0370269324002739}
}

@article{LHCb1,
  title = {Measurement of the forward $Z$ boson production cross-section in $pp$ collisions at $\sqrt{s}=7$ TeV},
  author = {Aaij, R. and others},
  collaboration = {LHCb Collaboration},
  journal = {JHEP},
  volume = {08},
  pages = {039},
  year = {2015},
  doi = {10.1007/JHEP08(2015)039}
}

@article{LHCb2,
  title = {Measurement of forward $W$ and $Z$ boson production in $pp$ collisions at $\sqrt{s}=8$ TeV},
  author = {Aaij, R. and others},
  collaboration = {LHCb Collaboration},
  journal = {JHEP},
  volume = {01},
  pages = {155},
  year = {2016},
  doi = {10.1007/JHEP01(2016)155}
}

@article{LHCb3,
  title = {Measurement of the forward $Z$ boson production cross-section in $pp$ collisions at $\sqrt{s}=13$ TeV},
  author = {Aaij, R. and others},
  collaboration = {LHCb Collaboration},
  journal = {JHEP},
  volume = {09},
  pages = {136},
  year = {2016},
  doi = {10.1007/JHEP09(2016)136}
}

@article{CMS1,
  title = {Measurement of the rapidity and transverse momentum distributions of $Z$ bosons in $pp$ collisions at $\sqrt{s}=7$ TeV},
  author = {Chatrchyan, S. and others},
  collaboration = {CMS Collaboration},
  journal = {Phys. Rev. D},
  volume = {85},
  issue = {3},
  pages = {032002},
  year = {2012},
  month = {Feb},
  publisher = {American Physical Society},
  doi = {10.1103/PhysRevD.85.032002},
  url = {https://link.aps.org/doi/10.1103/PhysRevD.85.032002}
}

@article{CMS2,
  title = {Measurements of differential $Z$ boson production cross sections in proton-proton collisions at $\sqrt{s}=8$ TeV},
  author = {Khachatryan, V. and others},
  collaboration = {CMS Collaboration},
  journal = {Phys. Rev. D},
  volume = {91},
  pages = {052008},
  year = {2015},
  doi = {10.1103/PhysRevD.91.052008}
}

@article{CMS3,
  title = {Measurements of differential $Z$ boson production cross sections in proton-proton collisions at $\sqrt{s}=13$ TeV},
  author = {Sirunyan, A. M. and others},
  collaboration = {CMS Collaboration},
  journal = {JHEP},
  volume = {12},
  pages = {061},
  year = {2019},
  doi = {10.1007/JHEP12(2019)061}
}

@article{ATLAS1,
  title = {Measurement of the $Z/\gamma^*$ boson transverse momentum distribution in $pp$ collisions at $\sqrt{s}=7$ TeV with the ATLAS detector},
  author = {Aad, G. and others},
  collaboration = {ATLAS Collaboration},
  journal = {JHEP},
  volume = {09},
  pages = {145},
  year = {2014},
  doi = {10.1007/JHEP09(2014)145}
}

@article{ATLAS2,
  author = "{ATLAS Collaboration}",
  title = "{Measurement of the transverse momentum and $\phi^*_{\eta}$ distributions of Drell--Yan lepton pairs in proton--proton collisions at $\sqrt{s}=8$ TeV with the ATLAS detector}",
  eprint = {1512.02192},
  archivePrefix = {arXiv},
  primaryClass = {hep-ex},
  reportNumber = {CERN-PH-EP-2015-275},
  doi = {10.1140/epjc/s10052-016-4070-4},
  journal = {Eur. Phys. J. C},
  volume = {76},
  number = {5},
  pages = {291},
  year = {2016}
}

@misc{ATLAS3,
  title = {Measurement of the transverse momentum distribution of Drell-Yan lepton pairs in proton-proton collisions at $\sqrt{s}=13$ TeV with the ATLAS detector},
  author = "{ATLAS Collaboration}",
  year = {2020},
  eprint = {1912.02844},
  archivePrefix = {arXiv},
  primaryClass = {hep-ex},
  url = {https://arxiv.org/abs/1912.02844}
}

@article{Catani_1996,
   title={The resummation of soft gluons in hadronic collisions},
   volume={478},
   ISSN={0550-3213},
   url={http://dx.doi.org/10.1016/0550-3213(96)00399-9},
   DOI={10.1016/0550-3213(96)00399-9},
   number={1-2},
   journal={Nuclear Physics B},
   publisher={Elsevier BV},
   author={Catani, S},
   year={1996},
   month=Oct, pages={273–310} }

@article{Kuleza_2002,
  title={Joint resummation in electroweak boson production},
  author={Kulesza, Anna and Sterman, George and Vogelsang, Werner},
  journal={Physical Review D},
  volume={66},
  number={1},
  pages={014011},
  year={2002},
  publisher={APS}
}

@article{Laenen_2000,
  title={Higher-order QCD corrections in prompt photon production},
  author={Laenen, Eric and Sterman, George and Vogelsang, Werner},
  journal={Physical Review Letters},
  volume={84},
  number={19},
  pages={4296},
  year={2000},
  publisher={APS}
}

@article{Kuleza_2004,
  title={Joint resummation for Higgs boson production},
  author={Kulesza, Anna and Sterman, George and Vogelsang, Werner},
  journal={Physical Review D},
  volume={69},
  number={1},
  pages={014012},
  year={2004},
  publisher={APS}
}

@article{DYTurbo_fit,
    author = "Camarda, Stefano and Ferrera, Giancarlo and Rossi, Lorenzo",
    title = "{Drell-Yan lepton pair production at low invariant masses: transverse-momentum resummation and non-perturbative effects in QCD}",
    eprint = "2508.06201",
    archivePrefix = "arXiv",
    primaryClass = "hep-ph",
    doi = "10.1007/JHEP01(2026)150",
    journal = "JHEP",
    volume = "01",
    pages = "150",
    year = "2026"
}

@misc{pareto_explorer,
  title        = {{Interactive Pareto Explorer for Symbolic TMD PDF Results}},
  year         = {2026},
  howpublished = {\url{https://wmdataphys.github.io/Results/Symbolic_TMD_PDF/pareto_explorer_global_maxexp_front.html}},
}

@article{Dotson:2025omi,
    author = "Dotson, Andrew and others",
    title = "{Generalized Parton Distributions from Symbolic Regression}",
    journal = "arXiv preprint",
    eprint = "2504.13289",
    archivePrefix = "arXiv",
    primaryClass = "hep-ph",
    year = "2025"
}

@article{Makke:2025zoy,
    author = "Makke, Nour and Chawla, Sanjay",
    title = "{Inferring interpretable models of fragmentation functions using symbolic regression}",
    eprint = "2501.07123",
    archivePrefix = "arXiv",
    primaryClass = "hep-ph",
    doi = "10.1088/2632-2153/adb3ec",
    journal = "Mach. Learn. Sci. Tech.",
    volume = "6",
    number = "2",
    pages = "025003",
    year = "2025"
}

@inproceedings{Morales-Alvarado:2024jrk,
    author = "Morales-Alvarado, Manuel and Conde, Daniel and Bendavid, Josh and Sanz, Veronica and Ubiali, Maria",
    title = "{Symbolic regression for precision LHC physics}",
    booktitle = "{Postponed: Machine Learning and the Physical Sciences}: {Workshop at NeurIPS 2024}",
    eprint = "2412.07839",
    archivePrefix = "arXiv",
    primaryClass = "hep-ph",
    year = "2024"
}

@article{Bendavid:2025urn,
    author = "Bendavid, Josh and Conde, Daniel and Morales-Alvarado, Manuel and Sanz, Veronica and Ubiali, Maria",
    title = "{Angular coefficients from interpretable machine learning with symbolic regression}",
    eprint = "2508.00989",
    archivePrefix = "arXiv",
    primaryClass = "hep-ph",
    doi = "10.1007/JHEP02(2026)081",
    journal = "JHEP",
    volume = "02",
    pages = "081",
    year = "2026"
}

@article{Kang:2026mod,
  author = {Kang, Zhong-Bo and Sellers, Luke and Zhang, Congyue and Zhou, Curtis},
  title = {{AI-assisted modeling and Bayesian inference of unpolarized quark transverse momentum distributions from Drell-Yan data}},
  eprint = {2604.14133},
  archivePrefix = {arXiv},
  primaryClass = {hep-ph},
  journal = {arXiv preprint},
  year = {2026},
}
